\documentclass[onecolumn,12pt]{IEEEtran}
\usepackage{cite}
\usepackage[pdftex]{graphicx}
\graphicspath{{../pdf/}{../jpeg/}}
\DeclareGraphicsExtensions{.pdf,.jpeg,.png}
\usepackage{amsmath, amssymb}

\usepackage{algorithm}
\usepackage{algpseudocode}

\usepackage{array}

\usepackage{lscape}
\usepackage{booktabs}
\usepackage{braket}
\usepackage{multirow}

\usepackage[flushleft]{threeparttable}

\usepackage{pifont}
\newtheorem{definition}{Definition}[section]

\newtheorem{theorem}{Theorem}[section]

\newtheorem{proof}{Proof}[section]
\usepackage{tikz}
\usetikzlibrary{mindmap,trees,positioning,shapes,calc,fadings,arrows,shapes.arrows,shadows,backgrounds, positioning}

\tikzset{lineProver/.style={thick, darkgray}}
\tikzset{lineVerifier/.style={thick, orange}}
\tikzset{lineMallory/.style={thick, red}}

\tikzset{lineProverMsg/.style={font=\small,thick, darkgray, draw,->}}
\tikzset{lineVerifierMsg/.style={font=\small,thick, orange,draw,->}}
\tikzset{lineBobMsg/.style={font=\small,thick,orange, draw,->}}
\tikzset{lineMalloryMsg/.style={font=\small,thick, red,draw,->}}

\tikzset{bubbleProver/.style={rectangle, draw=darkgray,rounded corners,align = flush center,minimum height=0.7cm,minimum width=1.5cm}}
\tikzset{bubbleVerifier/.style={rectangle, draw=orange,rounded corners,align = flush center,minimum height=0.7cm,minimum width=1.5cm}}
\tikzset{bubbleMallory/.style={rectangle, draw=red,rounded corners,fill=red,align = flush center,minimum height=0.7cm,minimum width=1.5cm}}

\tikzset{bubble/.style={rectangle, rounded corners,fill=#1,align = flush center,minimum height=0.7cm,minimum width=1.5cm}}
\tikzset{FunctionBubble/.style={draw, fill=blue!30, circle, node distance=2cm}}
\tikzset{DecisionBubble/.style={draw, fill=blue!30, diamond, text badly centered,node distance=2cm}}

\tikzset{ChannelBubble/.style={draw, fill=blue!20, rectangle,minimum height=3em, minimum width=6em}}
\tikzset{bubble/.style={font=\footnotesize,rectangle, rounded corners,fill=blue!20,align = flush center,minimum height=0.7cm,minimum width=1.5cm}}
\tikzfading [name=arrowfading, top color=transparent!0, bottom color=transparent!95]
\tikzset{arrowfill/.style={top color=gray!20, bottom color=gray, general shadow={fill=black, shadow yshift=-0.8ex, path fading=arrowfading}}}
\tikzset{arrowstyle/.style={draw=gray,arrowfill, single arrow,minimum height=#1, single arrow,single arrow head extend=.2cm,}}
\tikzstyle{pinstyle} = [pin edge={to-,thick,black}]

\begin{document}

\title{A Zero-Knowledge Proof of Knowledge for Subgroup Distance Problem}

\author{Cansu~Betin~Onur
\IEEEcompsocitemizethanks{
\IEEEcompsocthanksitem C. Betin Onur is partially support by NGI.}%
}

\maketitle

\begin{abstract}

In this study, we introduce a novel zero-knowledge identification scheme based on the hardness of the subgroup distance problem in the Hamming metric. The proposed protocol, named Subgroup Distance Zero Knowledge Proof (SDZKP), employs a cryptographically secure pseudorandom number generator to mask secrets and utilizes a Stern-type algorithm to ensure robust security properties. \end{abstract}

\begin{IEEEkeywords}
Zero-knowledge proofs, Subgroup Distance Problem, Hamming Distance
\end{IEEEkeywords}

\section{Introduction}\label{sec:introduction}

Zero-knowledge proofs (ZKPs) are cryptographic protocols that enable one party (the prover) to convince 
another party (the verifier) to the truth of a statement without disclosing any additional information beyond the 
 fact that the statement is true \cite{mohr2007survey, goldreich1991proofs, goldreich2002zero}. Since their inception, ZKPs have become a fundamental tool in cryptographic research and applications, providing a robust method for secure authentication and privacy-preserving computations. Various identification schemes leveraging ZKPs have been proposed, each relying on different hard mathematical problems to ensure security.
 One of the pioneering works in this area is Stern's identification scheme, introduced in 1993, which is based on the hardness of the Syndrome Decoding (SD) problem. Stern's protocol laid the groundwork for code-based cryptographic systems, renowned for their resilience to structural attacks and potential for quantum security through the Fiat-Shamir transform. Over the years, numerous enhancements and variations of Stern's protocol have been developed, focusing on improving efficiency and reducing soundness errors.

In this study, we introduce a zero-knowledge identification scheme that relies on the Subgroup Distance Problem (SDP) in the Hamming metric, a problem known for its computational hardness in various metrics and its NP-completeness under certain conditions. The proposed protocol, named Subgroup Distance Zero Knowledge Proof (SDZKP), differs from existing schemes by leveraging the SDP, thus providing a fresh approach to zero-knowledge identification. By converting the confidential information held by the prover into integer tuples and employing a Stern-type algorithm, we ensure that our protocol inherits the robustness and security features of code-based systems.

SDZKP differs from other identification schemes in the literature in terms of the mathematical hard problem on which it is based. However, the confidential information held by the prover is converted into an integer tuple and a Stern type algorithm is executed in SDZKP. For this reason, we place our study in association with code-based protocols. The SDZKP protocol is designed to be both secure and efficient, making use of a cryptographically secure pseudorandom number generator (CSPRNG) for masking secrets. The protocol follows a three-step challenge-response structure, ensuring that the prover can convince the verifier of their knowledge of a secret without revealing any information about it. Our security analysis demonstrates that SDZKP achieves perfect completeness, 3-special-soundness, and statistical zero-knowledge, providing strong guarantees against potential attacks.

In 1993, Stern introduced a code-based identification scheme based on the hardness of the Syndrome Decoding (SD) problem\cite{stern1993new}. His work became a foundational work on code-based cryptography. One major feature that makes Stern scheme attractive is its robustness to structural attacks. In other words, possible attacks appears only to the underlying hard problem. Using Fiat-Shamir transform, Stern protocol can be converted to a quantum secure signature scheme. 
 Various studies have been published focusing on improvement on efficiency and security of Stern protocol. These works are named as Stern-type protocols.  First improvement given by 
 V{\'e}ron \cite{veron1997improved} using General Syndrome Decoding problem and reduced the number of rounds required.
 Both protocols have $\frac{2}{3}$ soundness error. In 2011, Cayrel et. al. \cite{cayrel2010zero} and Aguilar et. al.  \cite{aguilar2011new} reduced Stern and V{\'e}ron protocols soundness error respectively up to $\frac{1}{2}$ using $5$-round protocols. In 2022, Bidoux et. al. \cite{bidoux2022quasi} introduced an adaptation of the protocol given by Aguilar, Gaborit and Schrek (AGS) \cite{aguilar2011new} on quasi-cyclic SD problem. Also in \cite{feneuil2023shared} a zero knowledge protocol achieving soundness error $\frac{1}{n}$ for arbitrary chosen $n$ is presented under the constrain that the verifier trust some of the variables sent by the prover.

This paper is organized as follows: Section \ref{sec:pre} covers the necessary preliminaries, including basic cryptographic definitions and an overview of the Subgroup Distance Problem. In Section \ref{sect:sdzkp}, we present the detailed design of the SDZKP protocol. Section \ref{sect:secanalysis} provides a comprehensive security analysis. Finally, we conclude the paper highlighting the contributions and potential future directions for research.

\section{Preliminaries}\label{sec:pre}

For any positive integer $m$ the set $\{1, \dots m  \}$ is denoted by $[m].$ The group of permutations on the set $[n]$ is denoted by $S_n.$ For a finite set $A,$ $ a \overset{{\scriptscriptstyle\$}}{\leftarrow} A$ denotes that $a$ is taken uniformly at random form $A.$
The term ``probabilistic polynomial time'' is abbreviated by PPT. On inputs $in_P, in_V$ respectively, the transcript of two parties $P$ and $V$ is denoted by $View(\langle P(in_P), V(in_V)\rangle)$ and an execution between $P$ and $V$ giving output $out$ is denoted by $\langle P(in_P),V(in_V)\rangle \rightarrow out.$
  As the subgroup distance problem is defined on the symmetric group $S_n,$ the security parameter of the given scheme is $n.$ 

\subsection{Subgroup Distance Problem (SDP)}\label{sect:sdp} 
Given a metric $d$ on the Symmetric group $ S_n .$ The distance of a permutation $\alpha \in S_n$ to a subgroup $H \leq S_n$ is defined as 
\[
d(\alpha, H)= \underset{h \in H}{min}   \ d(\alpha, h)
\]

\begin{definition}(\textbf{Subgroup Distance Problem (SDP)})
Given a set of elements $\lbrace g, h_1, \dots h_m \rbrace$ from $ S_n $ and given an integer $k.$ Decide whether the distance between $g$ and the subgroup $H=\langle h_1, \dots h_m \rangle$ is at most $k.$
    \end{definition}

We would like to draw readers attention to similarity of SDP problem and one of the closest vector problem (CVP) on integer lattices which is considered to be one of the hard problems for post-quantum cryptographic protocols. Roughly speaking, in both problems one is asked to decide (or search) existance of an element from a given subset close enough to given fixed element. 

 The computational complexity of SDP has been analyzed through various of metrics.  In 2006, Pinch showed that  SDP is NP-Complete\cite{pinch2007distance} with respect to Cayley Distance. Subsequently, in 2009, Buchheim et al. extended this result to other metrics such as Hamming, Kendall's tau, $l_p$, Lee's and Ulam's distances. Moreover, even when the subgroup is restricted to be an Abelian group of exponent two, the problem remains NP-complete across all these metrics.

The  Hamming distance of given two permutations $\alpha, \beta$ in $S_n$ is defined as the number of different entries of $\alpha$ and $\beta.$ 
\[
d(\alpha,\beta)=|\lbrace i | \ \alpha(i) \not= \beta(i)   \rbrace |.
 \]

\subsection{Basic Cryptographic Definitions}\label{sect:build}

\begin{definition}
A function $negl: \mathbb{N} \rightarrow \mathbb{R}_{\geq 0}$ is called negligible function if for any natural number $c$ there exists a lower bound $n_0$ such that $negl(n)<\frac{1}{n^c}$ for all $n \geq n_0.$     
\end{definition}

If the probability of a situation occurring is $1-negl(n)$ for some negligible function $negl,$ we say that the situation appears with overwhelming probability.

\begin{definition}
    For security parameter $n,$ two distribution ensembles ${D_n}, {E_n}$ are said to be 
    \textbf{computationally indistinguishable} 
    if for any PPT algorithm $A,$
    the value of the difference  $$ | \underset{x\leftarrow D_n}{Pr}[A(x)=1] - \underset{x \leftarrow E_n}{Pr}[A(x)=1] |$$ is a negligable function.

If the condition is true even if $A$ is allowed to be unbounded, then these distributions are said to be \textbf{statistically indistinguishable}. 

\end{definition}

Next we define commitment schemes which are  effective building blocks frequently used in ZKP design. A commitment scheme should satisfy three basic properties: correctness, binding and hiding properties. We give the definitions of these properties consequently after the definition of commitment schemes.

\begin{definition}
 A \textbf{commitment scheme} is a polynomial time algorithm triple $Com=(Setup,Commit, Ver)$  satisfying correctness, binding and hiding properties. The components of $Com$ are described as below:
 \begin{itemize}
     \item $Setup$: On input $1^n$, it outputs public parameters $PP$ determining 
    the message, the randomness, the commitment and the opening spaces. 
    Notation: $PP \leftarrow Setup(1^n).$
  \item $Commit$: On input $PP$ and a message $m,$ 
it outputs a commitment-opening pair formally represented as $(c,o) \leftarrow Commit(PP,m).$ 
            \item $Ver$: On input sequence $(PP,c,m,o)$ it outputs a bit $b.$ 
       The case $b=1$ refers to $c$ is a valid commitment for $m$ and $Ver$ accepts. In the case $b=0$, the committed value fails and $Ver$ rejects it.  Formally, we represent $Ver$ as $b\leftarrow Ver(PP,c,m,o).$
 \end{itemize}

\end{definition}

It is often that commitment algorithm triples take also some randomness $r$ as input. It is implicitly used in the above given definitions.

\begin{definition}
    A commitment scheme $Com=(Setup,Commit, Ver )$ satisfies the following properties:
    
\begin{itemize}
    \item \textbf{Correctness} if
\[
\operatorname{Pr}
\left[1\leftarrow Ver(PP,c,m,o)
:
(c,o) \leftarrow Commit(pp,m)
\right]=1
\]

\item  \textbf{Computationally (resp. Statistically) Binding} if there exists a negligible function $neg(n)$ such that for every PPT (resp. unbounded) algorithm $A$ below inequality holds:


\[
\operatorname{Pr}
\left[
\begin{array}{c}
pp \leftarrow Setup(1^n) \\
(c,m,m',o,o') \leftarrow A(pp)
\end{array}
:
\begin{array}{c}
m \neq m' \\
1\leftarrow Ver(PP,c,m,o)\\
1\leftarrow Ver(PP,c,m',o') \\
\end{array}
\right] \leq neg(n)
\]

    \item \textbf{Computationally (resp. Statistically) Hiding}  if for any two massages $m,m'$ and $pp \leftarrow Setup(1^n)$ the  distributions $Commit(pp,m)$ and $Commit(pp,m')$ 
 are computationally (resp. statistically) indistinguishable.
\end{itemize}   
\end{definition}

Next we define interactive protocols and zero-knowledge proof of knowledge. Informally, a zero-knowledge proof of knowledge is a two party protocol between a prover $P$ and a verifier $V$ such that  $P$ convinces $V$ that it has the knowledge of a desired secret without revealing any information about the secret. Let us put this in more symbolic language.

Let $R$ be an NP-relation and $L$ be the language corresponding to $R.$ That is $L = \{ x \ | \  \exists w : R(x, w) = 1 \}.$ In a zero-knowledge proof of knowledge protocol, for a common input value $x$, the prover $P$ convinces $V$ that it knows a witness $w$ such that $R(x, w) = 1.$

\begin{definition} [\textbf{Zero-knowledge proof of knowledge (ZKPoK)}]
A zero-knowledge proof of knowledge protocol for language $L$ with respect to a relation $R$ is a protocol between a pair of interactive machines $P$ and $V$ named prover and verifier where $P$ is computationally unbounded and $V$ is probabilistic polynomial-time. It should satisfy the following conditions:

\begin{itemize}
\item \textbf{Completeness: } For every $x \in L$, verifier $V$ always accepts after interacting with a prover $P$ having a witness $w.$

 $$Pr[\langle P(x,w), V(x)\rangle  \rightarrow 1] = 1$$

\item \textbf{Proof of Knowledge (with error $\epsilon$): } For every possibly cheating $T$-time PPT prover $P^{*}$ with $Pr[\langle P^{*},V(x)\rangle  \rightarrow 1] > \epsilon + e $ there exists a PPT algorithm $K$  (with running time polynomial in $\frac{1}{e} $ and $T$) such that; given rewindable black-box access to $P^{*},$ on input $x$ the algorithm $K$ outputs a $w'$ such that $R(x, w') = 1$ with success probability at least $\frac{1}{2}.$ Here $K$ is called the \textbf{knowledge extractor}.

\item \textbf{Zero Knowledge: } For every possibly cheating PPT verifier $V^{*}$, there exists a PPT algorithm $S,$ called \textbf{simulator}, such that on input $x$ it outputs a transcript $S(x)$ which is indistinguishable from $View(\langle P(x,w), V^{*}(x)\rangle ).$ The zero knowledge property is called computational, statistical or perfect zero knowledge depending on whether the two distributions are  computationally indistinguishable, statistical indistinguishable  or equal respectively.
\end{itemize}
\end{definition}

Under the restriction that $P$ is a PPT algorithm in above definition, the protocol refereed as "argument" instead of "proof."   i.e. we define zero knowledge argument of knowledge (ZKAoK).
A zero knowledge proof of knowledge is specified as \textbf{honest verifier zero knowledge proof(or argument) of knowledge} if the existence of a PPT simulator S giving output with indistinguishable distribution from $View(\langle P(x, w), V(x)\rangle )$ guarantied only for the \textit{honest} verifier $V.$

A stronger notion of knowledge soundness is (two)special-soundness. Here we give the definition of special-soundness for more generic case. Consider $3$-move protocols such that the rounds starts with the prover's move. The moves are named as commitment, challenge and response respectively. The transcripts are denoted by $(C,Ch,Rsp)$ in the sequel.

\begin{definition}
    A $3$-round protocol is said to have {\bf $k$-special-soundness} property if there exists a PPT algorithm $K$ such that for any given $k$ distinct excepted transcripts for the same commitment $C,$ say $(C,Ch_1,Rsp_1), \dots, (C,Ch_k,Rsp_k),$ the algorithm $K$ outputs a valid witness $w$. 
\end{definition} 

Under the assumption that $k=poly(x)$ for some polynomial, $k$-special-soundness strictly implies  knowledge soundness by a generic reduction with soundness error $\epsilon = (k - 1)/N$ , where $N$ is the cardinality of challenge space \cite{hazay2010efficient}. 

\begin{definition}
    A sigma ($\Sigma$) protocols is a $3$-round honest-verifier zero-knowledge proof of knowledge protocol satisfying $k$-special-soundness.
\end{definition}

\section{Subgroup Distance Zero Knowledge Proof (SDZKP)}\label{sect:sdzkp}

\begin{figure*}[ht]
\centering
 \begin{tikzpicture}[framed, auto, node distance=2.5cm,>=latex']
 \def \n {30}
 \def \height {15cm}
 \def \width {4cm}
 \def \Proverposx {0}
 \def \Proverposy {0}
 \def \Verifierposx {\width}
 \def \Verifierposy {0}

\foreach \s in {0,...,\n}
{
  \coordinate (a\s) at (\Proverposx, {\height - (\height / \n) * \s});
  \coordinate (b\s) at (\Verifierposx, {\height - (\height / \n) * \s});
};

\node[bubbleProver] (A) at (a0) {Prover};
\draw[lineProver] ($(A.south)$)--($(A.south)-(0,10)$);
\node[bubbleVerifier] (B) at (b0) {Verifier};
\draw[lineVerifier] ($(B.south)$)--($(B.south)-(0,10)$);

\node[bubble,left=1.5cm of a4]{

$ u \overset{{\scriptscriptstyle\$}}{\leftarrow} H,$\\ 
$ s \overset{{\scriptscriptstyle\$}}{\leftarrow} \mathbb{Z}$\\
 $uh, ug$\\ 
  $U=(uh(1) \dots uh(n) ) $;  \\
 $ G=(ug(1)  \dots ug(n)) $\\
 $R=vec_n(CSPRNG(s)) $\\
$Z_1=U+R,$ \ $Z_2=G+R,$  \\
  $C_1=Comm(Z_1)$; \\
  $C_2=Comm(Z_2)$; \\
  $C_3=Comm(s)$;\\
  $C=(C_{1}, C_{2}, C_{3})$
};
 \draw [lineProverMsg] (a5) -- node[above]{$C$} (b7);
 
\node[bubble,right=1cm of b7]{$Ch \overset{{\scriptscriptstyle\$}}{\leftarrow} \{0, 1, 2\}$};

\draw [lineVerifierMsg] (b8) -- node[above]{$Ch$} (a10);

 \node[bubble,left=1cm of a11]{If $Ch = 0$, $Rsp = \{Z_1, s\}$ \\
 If $Ch = 1$, $Rsp = \{Z_2, s\}$ \\
 If $Ch = 2$, $Rsp = \{Z_1, Z_2\}$};
 \draw [lineProverMsg] (a11) -- node[above]{$Rsp$} (b13);
 
 \node[bubble,right=0.5cm of b16]{

\begin{tabular}{ l } 

For $Ch=0$ obtain $R$, evaluate $U$\\
Check $uh \stackrel{?}{\in} H$, $C_{1} \stackrel{?}{=}  \text{Comm}(Z_1),$   
$C_3 \stackrel{?}{=}\text{Comm}(s)$ 

\\ \hline 
For $Ch=1$ obtain $R$, evaluate $u=(ug)g^{-1}$\\
Check $u\stackrel{?}{\in}H$, $C_{2} \stackrel{?}{=}  \text{Comm}(Z_2),$   
 $C_3 \stackrel{?}{=}\text{Comm}(s)$ 
 \\ \hline
For $Ch=2$, check $|\{ i \ | \ (Z_1-Z_2)_i \not= 0 \}|\stackrel{?}{\leq} k$, \\ 
 $C_{1} \stackrel{?}{=}  \text{Comm}(Z_1),$ 
$ C_{2} \stackrel{?}{=}  \text{Comm}(Z_2)$ 
\end{tabular}

};

\end{tikzpicture}
\caption{The message flow diagram depicting Subgroup Distance Zero Knowledge Proof (SDZKP).}
\label{fig:SDZKP}
\end{figure*}
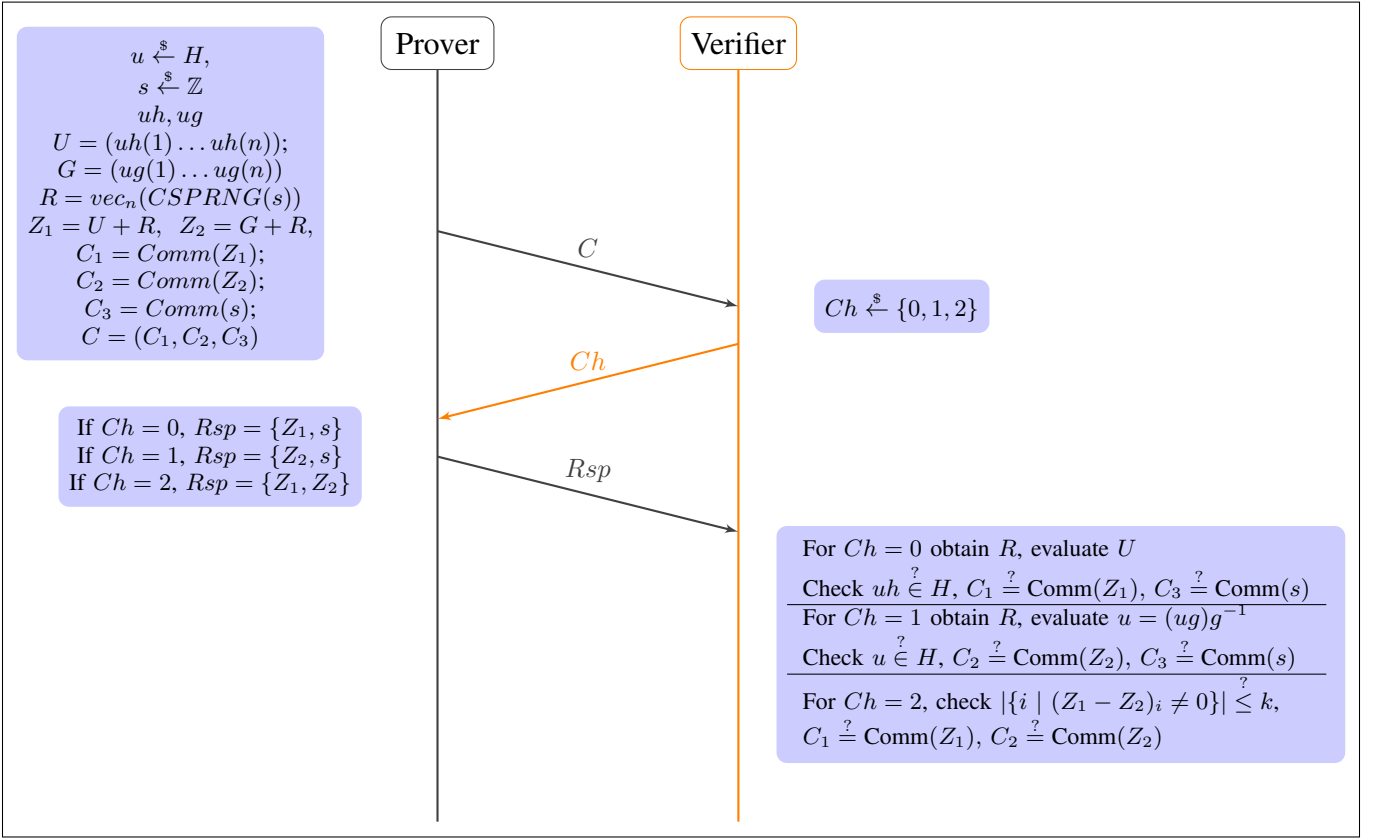

In this section, we present Subgroup Distance Zero Knowledge Proof (SDZKP) in steps and illustrate it in Figure \ref{fig:SDZKP}.

\subsection{Setup}\label{sect:setup}

In this section, under the assumption that the subgroup distance problem in Hamming metric is hard for parameters $k,n$ and the subgroup $H$ determined by the generators $h_1, \dots, h_m;$ we introduce a
zero knowledge identification scheme. 

The integers $k,n$ and a set of elements $\lbrace g, h_1, \dots h_m \rbrace$ from the Symmetric group $ S_n $ are assumed to be publicly known. The prover claims that it knows an element $h \in H=\langle h_1, \dots h_m \rangle$ such that $d(h,g)\leq k.$

\subsection{Protocol Design} \label{Pr1}

Employing subgroup distance problem we design a black-box statistical zero knowledge proof of knowledge protocol that we refer to as  Subgroup Distance Zero Knowledge Proof (SDZKP).
In this protocol, a cryptographically secure pseudorandom number generator (CSPRNG) is
 used for masking secrets.

\textbf{Step 1}: The prover selects an element $u\in H,$ and a seed integer $s$ for $CSPRNG$ uniform randomly.
 It generates length-$n$ integer tuples $U$ and $G$ where the 
 $i$th-entry of these tuples are $u(h(i))$ and $u(g(i)),$ respectively. Finally, it queries a random masking tuple $R=vec_n(CSPRNG(s)),$ evaluates and sends the commitments  $C_1=Comm(U+R)$,   $C_2=Comm( G+R),$ $C_3=Comm(s).$ Here addition $U+R$ and $G+R$ are component-wise addition of tuples. 

\textbf{Step 2}: The verifier generates a random challenge $Ch\in  \{0, 1, 2\}$ and sends it to the prover.

\textbf{Step 3}: Depending on challenge, the prover generates and sends a response $Rsp$ as follows:
\begin{itemize}
    \item If $Ch = 0$, $Rsp = \{Z_1, s\}$ where $Z_1=U+R$ 
    \item If $Ch = 1$, $Rsp = \{Z_2, s\}$ where $Z_2=G+R$ 
    \item If $Ch = 2$, $Rsp = \{Z_1, Z_2\}$
\end{itemize}

\textbf{Step 4}: This is the verification step. If all checks are valid, then verifier accepts. Otherwise, it rejects. 
\begin{itemize}
    \item For $Ch=0,$ the verifier first checks the validity of the commitments $C_{1} \stackrel{?}{=}  \text{Comm}(Z_1),$ $C_3\stackrel{?}{=}\text{Comm}(s).$ Then using the seed $s$, it obtains the tuple $R.$  It evaluates the tuple $U=Z_1-R$ and checks whether the corresponding permutation $uh$ is in $H.$ 

    \item For $Ch=1$, the verifier first checks the validity of the commitments $C_{2} \stackrel{?}{=}  \text{Comm}(Z_2)$ and $C_3\stackrel{?}{=}\text{Comm}(s).$  Then it obtains $R=vec_n(CSPRNG(s))$, computes $G=Z_2-R.$ It reconstructs the permutation $ug$ and evaluates $u=(ug)g^{-1}$. Then it checks whether $u\in H.$

    \item For $Ch=2$, the verifier checks commitments  
$C_{1} \stackrel{?}{=}  \text{Comm}(Z_1),$
$ C_{2} \stackrel{?}{=}  \text{Comm}(Z_2).$ Finally it evaluates the tuple $U-G$ and checks whether the number of  non-zero entries in $U-G$ is less than $k.$
\end{itemize}

\section{Security Analysis}
\label{sect:secanalysis}
In this section, we will prove that the given protocol is a  zero-knowledge proof system.

\begin{theorem}
Protocol SDZKP is a black-box statistical zero knowledge proof of knowledge protocol with knowledge soundness error $\frac{2}{3}.$
\end{theorem}
\begin{proof}
We will show completeness, $3$-special-soundness and statistical zero knowledge properties.
\begin{itemize}
    \item \textbf{Perfect Completeness:}
 If a prover P having knowledge of an element $h \in H$ with $d(h,g) \leq k$ follows the steps of the protocol, an honest verifier always accepts.
 It is known that the Hamming on permutation groups are left-invariant \cite{cameron2010complexity}. Therefore for any arbitrarily chosen $u \in H,$ $d(uh,ug)=d(h,g) \leq k.$ 
 Hence the proof of completeness is straightforward.

    \item \textbf{3-special-soundness:}
We show that under the assumption that the subgroup distance problem in Hamming metric is hard for parameters $k,n$ and for the subgroup $H$ determined by the generators $h_1, \dots, h_m$ and under the assumption that the used commitment function $Comm$ is computationally binding, SDZKP  is 3-special-sound.

We describe a knowledge extractor $K$ rewinding the protocol for the same randomness $u$ and $s$. We assume that $K$ gathers three excepted transcripts $(C, Ch_1, Rsp_0)$, $(C, Ch_2, Rsp_2)$,$(C, Ch_3, Rsp_2)$ from $P$ where $C=$ $(C_1,C_2,C_3)=$ $(Comm(Z_1),Comm(Z_2),Comm(s)).$ 
As there are only three choices for $b,$ without loss of generality we assume $Ch_1=0, Ch_2=1$ and $Ch_3=2.$ Then $Rsp_0 = \{Z_1^0, s^0\},$  $Rsp_1 = \{Z_2^1, s^1\},$  $Rsp_2 = \{Z_1^2, Z_2^2\}.$
 Commitment $C_1$ is verified for the cases $Ch=0,2$ and the commitment $C_2$ is verified for the cases $Ch=1,2.$ Then $Z_1^0=Z_1^2$ and $Z_2^1=Z_2^2$ by the binding property of $Comm.$ 
 Similarly $s^0=s^1$ guarantied by the validity of $C_3.$ Therefore we will express the values as $Z_1, Z_2, s$ without using the $Rsp$ index numbers.
  The extractor $K$ evaluates $U=Z_1-vec_n(CSPRNG(s))$ and $G=Z_2-vec_n(CSPRNG(s)).$ Then $K$ converts the sequences to permutations $ug$ and $uh.$ Lastly, it outputs $h= (ug[g^{-1}])^{-1}(uh).$
 
  The validity of first two transcripts guaranties that $uh \in H$ and  $u=ug[g^{-1}] \in H.$ Then, as $H$ is a group, we guarantee that $h=u^{-1}(uh)\in H.$ 

Also the last accepted transcript shows that the output $h$ satisfies the requred distance property $d(h,g)=d(uh,ug)= |\{ i \ | \ (Z_1-Z_2)_i \not= 0 \}|\ {\leq} k.$

   \item \textbf{Zero Knowledge:}
   Under the assumption that the used commitment scheme is statistically hiding, we show that the given protocol is statistically zero-knowledge by describing a simulator $S$ having black-box access to a malicious verifier $V^*.$
   The simulator $S$ given in Algorithm \ref{alg:sim1}  is build on challenge value
 prediction that will be chosen by $V^*$.  Obviously, $S$ runs in polynomial-time. It is allowed to rewind $V^*$ at most $M$ times. To see that $S$ generates a transcript statistically indistinguishable from the view of a real interaction, we present an alternative  simulator $S_0$ as an intermediate step in the discussion. The simulator $S_0$ has the knowledge of the secret $h$ and it follows the same stages with $S.$ i.e., it guesses the challenge value at stage 1 and it rewinds at stage 3 under the described situations in $S.$ The difference is, $S_0$ does not define or use a fake $h^*$ value. It evaluates commitments as described in $SDZKP$. 
Therefore, when $S_0$ makes a successful guess in $Ch,$ the distribution over the commitments viewed in $S_0$ and in a real proof is identical. At each attempt, it has success probability $\frac{5}{9}.$ 
That is $S_0$ outputs $\bot$ with at most probability $(\frac{4}{9})^M.$ Next we see that $S$ and $S_0$ gives statistically  indistinguishable outputs.
Algorithms $S$ and $S_0$ executes exactly in the same way except stage 2. At stage 2, While $S_0$  uses the knowlegde of the secret $h,$ $S$ uses a fake $h^*$ value. Both algorithm gives successful output only when $Ch,Ch^*\in \{0, 1\}$ or $Ch^*=2=Ch.$ 
In each case, the committed values are $Z_1=U+R,$  $Z_2=G+R$ and $s.$ 
In both algorithms $s$ is a uniform randomly chosen seed integer for known cryptographically secure pseudorandom number generator CSPRNG. So,  the distribution of chosen element $s$ and the generated integer string $R=vec_n(CSPRNG(s))$ in $S$ is the same as the distribution of $s$ and $R$ obtained in $S_0,$ respectively.  In both algorithms $U$ and $G$ are random shuffles of length-n integer the tuple $(1 2 \dots n).$ 
Therefore the distribution of the evaluated tuples $Z_1=U+R,$  $Z_2=G+R$ in $S$ and $S_0$ are statistically indistinguishable.

\begin{algorithm}
\caption{The algorithm of simulator $S$ for SDZKP.}
\label{alg:sim1}
\begin{algorithmic}[1]
\Require Public parameters and security parameter $M.$
\State  \textbf{STAGE 0}: 
\State $m=1$
\State  \textbf{STAGE 1}: Fix a random challange and a fake secret $h^*$ 
\State $Ch^* \overset{{\scriptscriptstyle\$}}{\leftarrow} \{0, 1, 2\}$ 

\If{$Ch^* \in \{0, 1\} $ }
    \State $h^* \overset{{\scriptscriptstyle\$}}{\gets} H $ 
\Else \Comment{$Ch^*=2$}
    \State $h^* {\gets} S_n$ such that $d(h^*,g){\leq} k$ 
    \Comment{can be done efficiently by manipulating $g.$}
\EndIf
\State  \textbf{STAGE 2}: Evaluate commitments using $h^*.$ 
 
\State $ s \overset{{\scriptscriptstyle\$}}{\leftarrow} \mathbb{Z}$
\State $U=(h^*(1) \dots h^*(n))$  
\State $G=(h^*g(1) \dots h^*g(n))$
\State $R=vec_n(CSPRNG(s))$
\State $Z_1=U+R$ 
\State $Z_2=G+R$  
\State $C_1=Comm(Z_1)$
\State $C_2=Comm(Z_2)$
\State $C_3=Comm(s)$
\State $C=(C_{1}, C_{2}, C_{3})$
\State  \textbf{STAGE 3}: Oracle access query to $V^*$ and obtain $Ch.$ 
\If{$Ch,Ch^*\in \{0, 1\}$ or $Ch^*=2=Ch$ }
    \State pass to Stage $4$
    
    \Else
    \State $m=m+1.$ 
    \If{$m\not=M,$}
    \State \textbf{goto} STAGE 1
    \Else 
    \State \textbf{return} $\bot$ and abort
\EndIf
\EndIf
\State \textbf{STAGE 4}: Follow the protocol response stage with $h^*$ 

\If{$Ch=0$ }
    \State  $Rsp = \{Z_1, s\}$     
\ElsIf{$Ch=1$}
    \State  $Rsp = \{Z_2, s\}$
\ElsIf{$Ch=2$}
\State $Rsp = \{Z_1, Z_2\}$
\EndIf

\\
\Return ($C$, $Ch$, $Rsp$)

\end{algorithmic}
\end{algorithm}

\end{itemize}
\end{proof}

\section{Conclusion}
\label{sect:conc}

In this paper, we have introduced a novel zero-knowledge identification scheme based on the Subgroup Distance Problem (SDP) in the Hamming metric, named Subgroup Distance Zero Knowledge Proof (SDZKP). Our protocol leverages the inherent computational hardness of the SDP to ensure robust security properties while maintaining efficiency. By utilizing a cryptographically secure pseudorandom number generator (CSPRNG) and a Stern-type algorithm, the SDZKP protocol achieves perfect completeness, 3-special-soundness, and statistical zero-knowledge proof of knowledge, making it resilient against adversaries.
 
Through this work, we contribute to the ongoing research in zero-knowledge proofs by presenting a new identification scheme that expands the toolkit available to cryptographers and security practitioners. The use of the Subgroup Distance Problem as the underlying hard problem opens new avenues for designing secure cryptographic protocols.

Future work may explore further optimizations of the SDZKP protocol, as well as its application in various cryptographic settings. Additionally, investigating the integration of SDZKP with other cryptographic primitives and protocols could provide new insights and advancements in the field of secure authentication and privacy-preserving computations.

\section*{Acknowledgement}
This work is partially supported by the NLnet foundation under the MoU number 2021-12-510.

\bibliographystyle{IEEEtranS}               
\bibliography{references.bib}

\end{document}